\documentclass[11pt,a4paper]{article}

\usepackage{jcappub}

\usepackage{subfig}

\usepackage[capitalize]{cleveref}
\usepackage{physics}

\newcommand{\lnd}{\mathcal{N}}
\newcommand{\lpf}{\mathcal{P}}
\newcommand{\lff}{\mathcal{B}}
\newcommand{\lcf}{\mathcal{A}}
\newcommand{\tsqrt}{b}

\title{Breaking the Strings: the signatures of Cosmic String Loop Fragmentation}

\author{Pierre Auclair}
\affiliation{Cosmology, Universe and Relativity at Louvain (CURL), Institute of Mathematics and
	Physics, University of Louvain, 2 Chemin du Cyclotron, 1348 Louvain-la-Neuve, Belgium}

\emailAdd{pierre.auclair@uclouvain.be}

\abstract{
    We study the impact of fragmentation on the cosmic string loop number density, using an approach inspired by the three-scale model and a Boltzmann equation.
    We build a new formulation designed to be more amenable to numerical resolution and present two complementary numerical methods to obtain the full loop distribution including the effect of fragmentation and gravitational radiation.
    We show that fragmentation generically predicts a decay of the loop number density on large scales and a deviation from a pure power-law.
    We expect fragmentation to be crucial for the calibration of loop distribution models.
}


\begin{document}

\maketitle
\flushbottom

\section{Introduction}

Cosmic strings are line-like topological defects that may have formed after a symmetry-breaking phase transition in the early Universe~\cite{Nielsen:1973cs,Kibble:1976sj} (see Refs.~\cite{Hindmarsh:1994re, Vilenkin:2000jqa, Vachaspati:2015cma} for reviews).
If present, cosmic strings are expected to be organized as a network of cosmological size, characterized by a set of length-scales growing as a constant fraction of the Hubble radius~\cite{Bennett:1987vf,Copeland:1991kz,Ringeval:2005kr}.

Such a long-lived network of cosmic strings would leave a number of imprints on cosmological observables depending on the adimensional string tension $G\mu/c^2$.
Gravitational wave observations in particular have renewed the interest in the field and may provide a new opportunity to detect cosmic strings and consequently probe the physics of the early Universe~\cite{LIGOScientific:2021nrg,EuropeanPulsarTimingArray:2023lqe,EPTA:2023xxk,NANOGrav:2023hvm,LISACosmologyWorkingGroup:2022jok,LISA:2024hlh,Abac:2025saz}.

Across the past decades, several models have been proposed to predict the properties of cosmic string networks embedded in an expanding cosmological background.
State-of-the-art models combine numerical simulations -- to probe the dynamics of the large non-linear scales --
and analytical modelling, necessary to make predictions on the smallest scales at which gravitational radiation~\cite{Polchinski:2007rg}, particle production~\cite{Matsunami:2019fss, Auclair:2019jip, Auclair:2021jud, Hindmarsh:2021mnl} and gravitational backreaction~\cite{Blanco-Pillado:2018ael, Blanco-Pillado:2019nto, Wachter:2024aos} dominate.

Such models include the LRS model~\cite{Lorenz:2010sm} in which the authors assume that loops are produced at all scales following a power-law Polchinski-Rocha loop production function~\cite{Polchinski:2006ee, Polchinski:2007rg, Dubath:2007mf}.
This model is calibrated against the loop number density found in the Nambu-Goto simulations of Refs.~\cite{Bennett:1987vf,Ringeval:2005kr}, see also \cite{Auclair:2019zoz, Auclair:2020oww} for more details on the procedure.
Later, the BOS model was proposed in Ref.~\cite{Blanco-Pillado:2013qja, Blanco-Pillado:2017oxo}.
There, the authors measure the loop production of stable loops from Nambu-Goto simulations~\cite{Blanco-Pillado:2010kol, Blanco-Pillado:2011egf}, and use it to reconstruct the loop number density, which the authors then compared with their numerical simulations in a later publication~\cite{Blanco-Pillado:2019tbi}.

Although of similar amplitude on large scales, these two models differ significantly on smaller scales, still inaccessible to numerical simulations.
It has been argued by the authors of the BOS model that the LRS model may violate energy conservation between the power received by infinite strings and the energy stored as loops~\cite{Blanco-Pillado:2019vcs}.
This claim relies on a number of assumptions, notably that the loops produced by the network become stable instantaneously.
It has been shown in Appendix A of Ref.~\cite{Auclair:2021jud} that including a fragmentation cascade can break the energy conservation argument by a factor of a decade.

The loop number density has also been explored extensively using field-theory simulations~\cite{Hindmarsh:2017qff, Hindmarsh:2018wkp, Figueroa:2020lvo, Baeza-Ballesteros:2024otj} and yield yet another prediction for the population of loops.
Abelian Higgs simulation of local strings show a rapid decay of loops through particle emission~\cite{Matsunami:2019fss,Hindmarsh:2021mnl, Blanco-Pillado:2023sap}, thus reducing their observational footprint.
Notably, the emission of particles from field-theory loops appears to depend dramatically on the cosmic string loop shape, loops having a simpler structure\footnote{The authors of Refs.~\cite{Hindmarsh:2021mnl,Baeza-Ballesteros:2024otj} considered square ``artificial'' loops for which the lifetime grows quadratically with length, as opposed to linearly for other types of loops.} surviving for longer and thus emitting more gravitational waves~\cite{Hindmarsh:2021mnl,Baeza-Ballesteros:2024otj}. 
One possible scenario to reconcile Nambu-Goto and Abelian-Higgs simulations is to interpret this energy lost to particles as a fragmentation cascade transporting the energy of the largest and more complex loops to simpler loops, smaller than the grid's resolution.

The objective of this work is to explore the consequences of fragmentation cascades on the loop distribution, a phenomenon that is most often neglected, including in both the LRS and BOS models\footnote{The BOS and LRS models do not include fragmentation \emph{explicitely}, but fragmentation does have an \emph{implicit} impact on the normalization of the loop production function.}
The main argument of this article is that fragmentation is a key and unavoidable property of cosmic string networks that may help reconciling decade-long disagrements in the cosmic string community, between field-theory and Nambu-Goto simulations on one side, and between Nambu-Goto simulations themselves on the other side.
This would be invaluable to GW observations whose constraints and predictions for the string tension $G\mu$ currently span several orders of magnitude due to theoretical uncertainties~\cite{LIGOScientific:2017ikf,Auclair:2019wcv,Auclair:2020oww,LIGOScientific:2021nrg,LISACosmologyWorkingGroup:2022jok,EPTA:2023xxk,Dimitriou:2025bvq}.

Scale-independent stochastic models of fragmentation cascades were proposed in Ref.~\cite{Smith:1987pu} but were found to be incompatible with numerical simulations.
Pioneering work on loop fragmentation in Minkowski background ~\cite{Sakellariadou:1987kc,Scherrer:1989ha,York:1989kf,Siemens:1994ir, Casper:1995ub, Copi:2010jw, Pazouli:2021orp} have shown that realistic loops, having a large number of harmonics, tend to fragment and cascade into a large population of stable and approximately planar loops. 
The question remains open in an expanding background, where one may expect a typical fragmentation length scaling with time to appear.
It should be noted that the first generations of Nambu-Goto simulations already highlighted the importance of fragmentation to determine the properties of the networks even on large scales~\cite{Bennett:1987vf,Bennett:1989yp,Sakellariadou:1990nd}, thus leading to some of the first models of cosmic string population with elements of fragmentation~\cite{Bennett:1985qt,Bennett:1986zn}.
The importance of small-scale structure on cosmic strings was also discussed in Ref.~\cite{Copeland:2009dk} through the study of the sharpness distribution of kinks.
In this work, we resurrect and adapt the model of Refs.~\cite{Steer:1998qg,Magueijo:1999qc} describing the dynamics of a network of interacting loops in an expanding cosmological background.
This model was inspired by the three-scale model of Refs.~\cite{Austin:1993rg, Copeland:1998na}.

In this article, we propose to adapt the model of Ref.~\cite{Steer:1998qg} to study loop fragmentation specifically, in a format that is more amenable to numerical resolution.
We present in \cref{sec:model} the assumptions underlying our fragmentation model, what were the features locking further progress in this direction and how we propose to circumvent them.
In \cref{sec:methods}, we present two complementary numerical methods to solve our fragmentation model.
The Unconnected-Loop Model (ULM) of \cref{sec:ulm} takes a ``micro'' approach by simulating large samples of random fragmentation cascades.
To the contrary, the Integro-Differential Equation (IDE) of \cref{sec:ide} takes a ``macro'' approach by solving the Boltzmann equation for the loop distribution.
Results for the loop distribution in the radiation and matter era are presented and discussed in \cref{sec:discussion}. 

\section{Fragmentation model}
\label{sec:model}

\subsection{Model and assumptions}

The model of Refs.~\cite{Steer:1998qg,Magueijo:1999qc} describes a network of interacting loops in an expanding background.
For example, two loops of invariant lengths $\ell_1, \ell_2$ may collide to form a larger loop of size $\ell_1 + \ell_2$ with a rate
\begin{equation}
    \lcf(\ell_1, \ell_2; \ell_1 + \ell_2) = \tilde{\chi} \ell_1 \ell_2 \, ,
\end{equation}
where $\tilde{\chi}$ is homogeneous to a velocity and assuming that two loops have cross-sections proportional to their lengths.
To the contrary, loops of length $\ell$ may fragment through self-intersection into two loops of smaller sizes $y < \ell - y$ with a rate given by the loop fragmentation function $\lff(y, \ell - y; \ell)$.
The actual form of the fragmentation function is still unknown, but some ansatz were proposed in Refs.~\cite{Copeland:1998na, Steer:1998qg} using arguments of detailed balance.

In model of Refs.~\cite{Steer:1998qg,Magueijo:1999qc}, the loops were assumed to lose energy to gravitational waves (GWs) and to stretch with the expansion on the Universe.
Under these assumptions, the authors wrote a non-linear Boltzmann equation for the loop distribution.
The authors were able to obtain the asymptotic behaviour of the loop distribution in certain scenarios but a complete analytical or numerical resolution remained an open challenge.

In this article, we propose a new simplified version of this model, designed specifically to study loop fragmentation and provide two independent resolution methods.
Our model keeps some aspects of these previous works while performing a number of simplifications.

Contrary to Refs.~\cite{Steer:1998qg,Magueijo:1999qc}, we restrict our domain to sub-Hubble loops and therefore neglect the effect of expansion on the loops.
Instead, we assume that a scaling infinite string network exists and interacts with the loop population through a boundary term, i.e.\ a one-scale loop production function
\begin{equation}
    \lpf(\ell, t) = C t^{-5} \delta\qty(\frac{\ell}t - \alpha)\, ,
    \label{eq:lpf}
\end{equation}
in which $C$ is a constant of order unity and $\alpha = \order{0.1}$ is the limit between the sub-Hubble and super-Hubble loops.

Another simplification in our model is that we also neglect the possible collisions between loops thus setting $\lcf = 0$.
This approximation is perhaps the key ingredient that makes our numerical resolutions tractable, so let us emphasize why we believe this approximation makes physical sense.
First, we find in Nambu-Goto simulations that collisions are rare events compared to fragmentations, and a significant fraction of these collisions are actually reconnections, i.e.\ a loop fragments into two smaller loops that reconnect after some time. 
Therefore, one could include reconnections through a redefinition of the fragmentation function.
Collisions between independent loops are rare events that we chose to neglect.
Neglecting the collision term makes the non-linear Boltzmann equation \emph{linear} and an \emph{initial value problem} more suited to numerical resolution.

\begin{figure}
    \includegraphics[width=\textwidth]{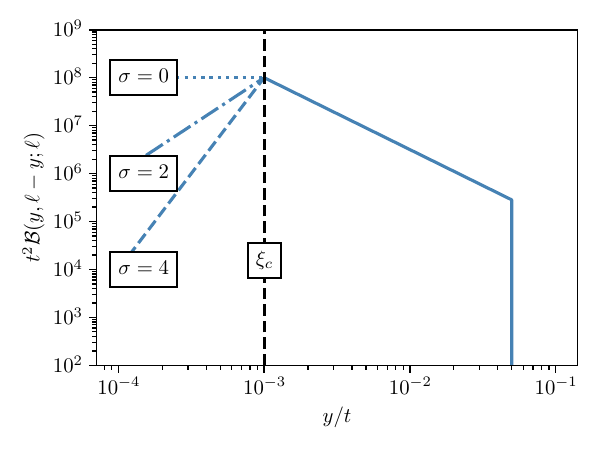}
    \caption{The loop fragmentation function for a parent loop $\ell / t = 0.1$ and different assumptions for the small-scale behaviour $\sigma$.}
    \label{fig:sigma}
\end{figure}
Finally, we assume the existence of a scaling correlation length $\xi = \xi_c t$.
Above this typical correlation length $\xi$, cosmic strings can be viewed as random walks with fragments of length $\xi$.
Therefore, the loop fragmentation function above the correlation length can be written as~\cite{Austin:1993rg}
\begin{equation}
    \lff(y > \xi, \ell - y; \ell) = \frac{\chi \ell}{(\xi y)^{3/2}} 
    \label{eq:lff-simple}
\end{equation}
with $\chi = \order{0.2}$ homogeneous to a velocity.
As mentioned in the introduction, various ansatz for the loop fragmentation function on all scales were proposed using arguments of detailed balance~\cite{Copeland:1998na,Steer:1998qg}.
To keep things simple, we introduce a parameter $\sigma$ to capture our lack of knowledge on the precise form of the loop fragmentation function
\begin{equation}
    \lff(y, \ell - y; \ell) = \frac{\chi \ell}{(\xi y)^{3/2}} \Theta(y - \xi) + \frac{\chi \ell}{\xi^3} \left(\frac{y}{\xi}\right)^{\sigma} \Theta(\xi - y) \, .
    \label{eq:lff-full}
\end{equation}
The impact of $\sigma$ is illustrated in \cref{fig:sigma}, and we keep $\sigma$ as a free parameter in the rest of the article.
Unless specified otherwise, the equations in the main text adopt the limit $\sigma \to \infty$, i.e.\ fragmentation is forbidden below the correlation length $\xi$, for simplicity.
The full expressions are given in \cref{app:sigma}.

\subsection{Master equation}

We rewrite the rate equation of Ref.~\cite{Steer:1998qg} for the number density of sub-Hubble loops under our new assumptions and obtain a linear partial differential and Integro-Differential Equation (IDE)
\begin{equation}
    \begin{split}
        \pdv{t} \qty[a^3(t) \lnd] + \pdv{\ell} \qty[\dot{\ell} a^3(t) \lnd] =
        a^3(t) \lpf(t, \ell)
        - a^3(t) \lnd(t, \ell) \int_\xi^{\ell/2} \lff(y, \ell-y ;\ell) \dd{y} \\
        + a^3(t) \int_{2\ell}^{\alpha t} \lff(\ell, L - \ell; L) \lnd(t, L) \dd{L}
        + a^3(t) \int_\ell^{2\ell} \lff(L-\ell, \ell; L) \lnd(t, L) \dd{L} \, .
    \end{split}
    \label{eq:master}
\end{equation}
The left-hand side of this rate equation encodes the conservation of the loop number density in an expanding background with scale factor $a(t)$
\begin{equation*}
    \pdv{t} \qty[a^3(t) \lnd] + \pdv{\ell} \qty[\dot{\ell} a^3(t) \lnd] \,,
\end{equation*}
in which the $\dot{\ell}$ term accounts for energy losses during the lifetime of the loop, for example through the production of GWs $\dot{\ell} = -\Gamma G \mu$~\cite{Vachaspati:1984gt,Allen:1991bk}.
In general, we assume that the universe is dominated by a single specie, therefore $a(t) = t^\nu$, with $\nu=1/2$ in radiation era and $\nu=2/3$ in matter era.

The right-hand side contains, from left to right, the loop production function $\lpf$ of \cref{eq:lpf}, which can be interpreted as a boundary term.
Then, the integral of the loop fragmentation function from \cref{eq:lff-full} gives the probability that a loop of length $\ell$ decays through fragmentation
\begin{equation*}
    - a^3(t) \lnd(t, \ell) \int_\xi^{\ell/2} \lff(y, \ell-y ;\ell) \dd{y} \, ,
\end{equation*}
forming two children loops of lengths $y$ and $\ell - y$. 
Eventually, these two children loops are included back into the loop distribution by the last two terms
\begin{equation*}
    a^3(t) \int_{2\ell}^{\alpha t} \lff(\ell, L - \ell; L) \lnd(t, L) \dd{L}
    + a^3(t) \int_\ell^{2\ell} \lff(L-\ell, \ell; L) \lnd(t, L) \dd{L} \, .
\end{equation*}

Note that this IDE is manifestly linear now that we have neglected collisions.
Moreover, \cref{eq:master} can be recast into a system using ordinary derivatives and integrals with scaling variables $\gamma = \ell / t$ and $\lnd(t, \ell) = t^{-4} n(\gamma)$
\begin{equation}
    \begin{split}
        \Gamma G \mu \dv{n}{\gamma}
    - (3\nu - 4) n(\gamma)
    - n(\gamma) \int_\xi^{\gamma / 2} \frac{\chi \gamma}{(\xi_c z)^{3/2}} \dd{z}
    = -C \delta(\gamma - \alpha) \\
    - \int_{2\gamma}^\alpha \frac{\chi}{(\xi_c \gamma)^{3/2}} n(Z) \dd{Z}
    - \int_{\gamma}^{2\gamma} \frac{\chi Z}{[\xi_c (Z - \gamma)]^{3/2}} n(Z) \dd{Z} \, .
    \end{split}
    \label{eq:master-scaling}
\end{equation}
We moved the self-intersection on the left-hand side to show that this term is effectively a decay term and sets the lifetime of the sub-Hubble loops.

Note that \cref{eq:master-scaling} is an \emph{initial-value problem}, or ``\emph{triangular}'' in the sense that the production of loops by the infinite string network sets the boundary condition at $\gamma = \alpha$.
Then, for any $\gamma \in ]0, \alpha[$, the right-hand side of the IDE only requires values of $n(Z)$ for $Z > \gamma$.
If we discretized this equation, one could write this system as a triangular square matrix with zero coefficients below the diagonal.

\section{Numerical methods}
\label{sec:methods}

In this section, we propose two independent methods to solve either the master equation in physical units \cref{eq:master} or the master equation in scaling units \cref{eq:master-scaling}.

Let us first explain the numerical difficulties that appear when trying to solve this system with traditional IDE solvers such as the python package IDESOLVER~\cite{Karpel2018}.
Traditional IDE solvers first solve the left-hand side -- or homogeneous part -- of \cref{eq:master-scaling} $n_0(\gamma)$.
Then, they inject $n_0(\gamma)$ in the integrals of the source term and find $n_1(\gamma)$, and continue this iterative process until convergence.
In our model, this is equivalent to computing the loop distributions of each of the nth generation loops in a cascade of potentially hundreds or thousands of generations.
This process is highly ineffective and convergence becomes exponentially more difficult as we increase the parameters driving fragmentation, i.e.\ $\chi \to 1$ and $\xi_c \ll 1$.

As a consequence, we designed two techniques specifically to circumvent these difficulties.
In \cref{sec:ulm}, we present a technique that reconstructs the loop number density bruteforcely by simulating a large sample of fragmentation cascades.
Then in \cref{sec:ide}, we present another technique that takes advantage of the ``triangular'' nature of \cref{eq:master-scaling}.

\subsection{Unconnected Loop Model (ULM)}
\label{sec:ulm}

\begin{figure}
    \centering
    \includegraphics[height=.4\textheight]{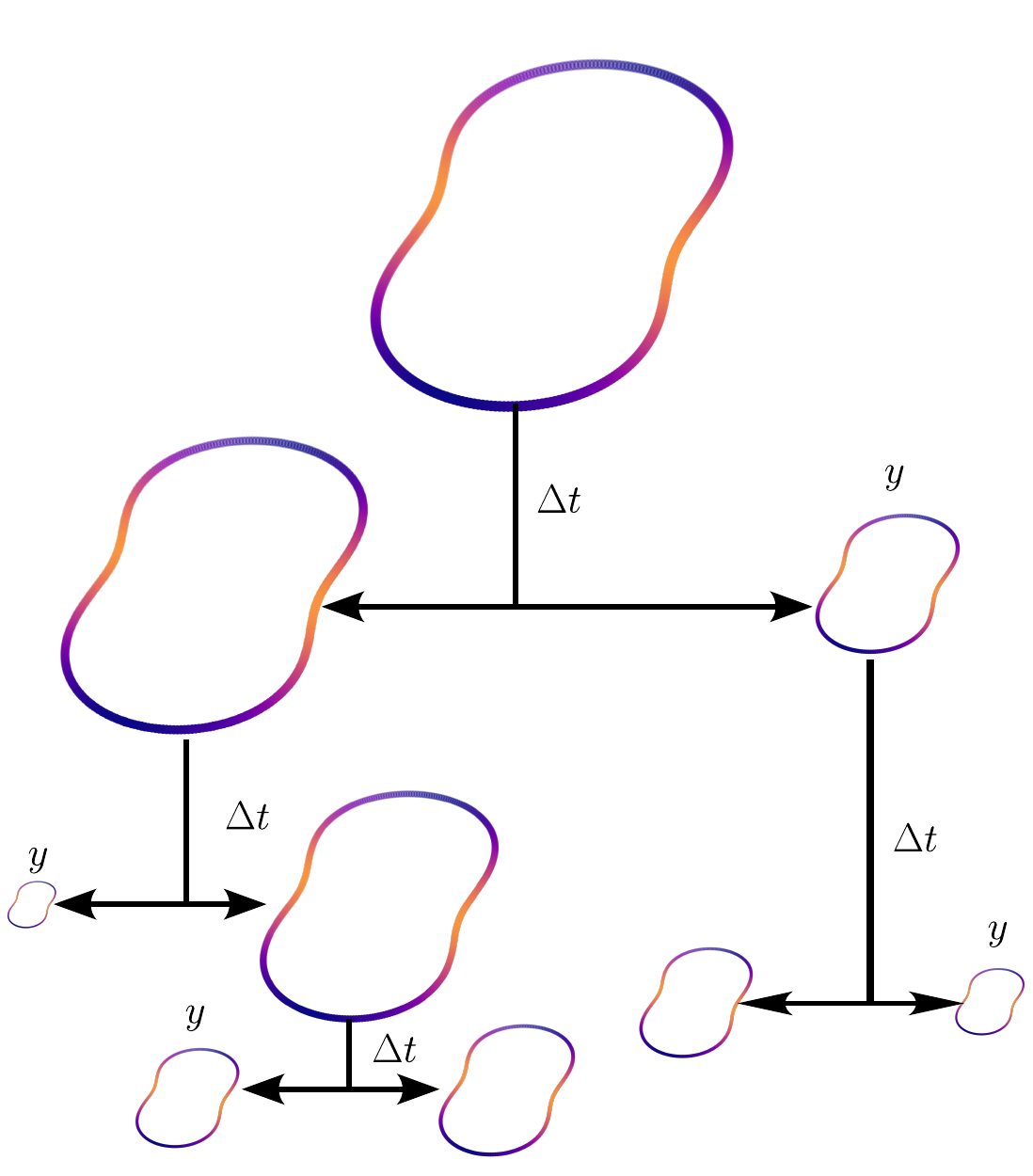}
    \caption{
        Schematic view of a loop fragmentation cascade. 
        From an initial loop, we draw its lifetime $\Delta t$ randomly and fragment it into two children loops.
        The sizes of the children loops are also drawn randomly from the loop fragmentation function $\lff(y, \ell- y; \ell)$.
    }
    \label{fig:cascade}
\end{figure}

\begin{figure}
    \includegraphics[width=\textwidth]{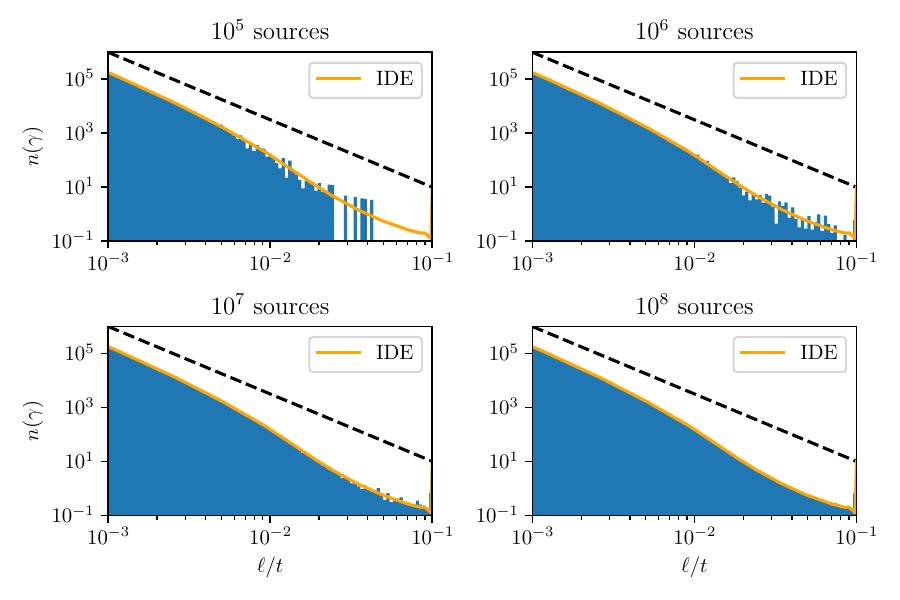}
    \caption{
        Comparison between the two numerical methods, the Unconnected Loop Model (ULM) in blue and the custom IDE solver in yellow, as we increase the number of fragmentation cascades. 
        The parameters for this figure are $(C, \alpha, \chi, \xi_c, \sigma) = (1, 0.1, 0.2, 10^{-2}, 0)$ in radiation era.}
    \label{fig:comparison}
\end{figure}

The Unconnected Loop Model (ULM) is designed to simulate a large sample of fragmentation cascades in physical units following \cref{eq:master}.
Note that it does not assume scaling apart from the boundary at $\alpha$ and the correlation length $\xi=\xi_c t$.
We neglect the gravitational radiation in this part $\Gamma G \mu = 0$.

Schematically, we give ourselves a minimum time $t_\mathrm{min}$ and an observation time $t_\mathrm{obs}$.
Then, we sample individual fragmentation cascades by iterating the following procedure, as illustrated in \cref{fig:cascade}.

First, we sample a random loop produced by the loop production function in a time $[t_\mathrm{min}, t_\mathrm{obs}]$.
In practice, we compute the cumulative probability to form a loop in a time range $[t_\mathrm{min}, t]$, in a volume which was smaller by a factor of $a^3(t)$ at the time of formation
\begin{equation}
    \begin{split}
        F(t) =
        \frac{\int_0^\infty \dd{\ell}\int_{t_\mathrm{min}}^t C \delta(\ell / t - \alpha) t^{3 \nu - 5} \dd{t'}}{\int_0^\infty \dd{\ell} \int_{t_\mathrm{min}}^{t_\mathrm{obs}} C \delta(\ell / t - \alpha) t^{3 \nu - 5} \dd{t'}}
        = \frac{t^{3 \nu - 3} - t_\mathrm{min}^{3\nu - 3}}{t_\mathrm{obs}^{3 \nu - 3} - t_\mathrm{min}^{3\nu - 3}} \, .
    \end{split}
\end{equation}
To draw the formation time $T$ -- and therefore the loop size $\alpha T$ -- we use a random variable $X \in \mathcal{U}(0, 1)$ with uniform distribution
\begin{equation}
    T = F^{-1}(X)
    = \qty[t_\mathrm{min}^{3\nu - 3} + X \qty(t_\mathrm{obs}^{3 \nu - 3} - t_\mathrm{min}^{3\nu - 3})]^{1/(3\nu - 3)} \, .
\end{equation}

As we saw in \cref{eq:master}, the fragmentation term can be interpreted as a characteristic decay time set by
\begin{equation}
    \tau^{-1} = \int_\xi^{\ell/2} \lff(y, ,\ell - y; \ell) \dd{y}
    = \frac{2 \chi \ell}{(\xi_c t)^{3/2}} \qty[(\xi_c t)^{-1/2} - \qty(\frac{\ell}2)^{-1/2}] \, .
\end{equation}
As a consequence, the probability that a loop survives until time $t$, given that the loop was formed at time $t_\star$ is given by the cumulative probability distribution
\begin{equation}
    G(t) = \exp[-\int_{t_\star}^t \tau^{-1} \dd{t}]\, .
    \label{eq:g}
\end{equation}
Using the same technique as for the formation time, we can draw the fragmentation time $\Delta t$ of the loop using a uniform variable $X \in \mathcal{U}(0, 1)$
\begin{equation}
    \Delta t = G^{-1}(X).
\end{equation}
In practice, one has to find the roots of the following quadratic equation. Using the notation $\tsqrt=t^{-1/2}$
\begin{equation}
    \tsqrt^2 - \tsqrt_\star^2 - 2\sqrt{\frac{2\xi_c}{\ell}} (\tsqrt - \tsqrt_\star) = \frac{\xi_c^2}{2\chi\ell}\ln(X) \, .
    \label{eq:lifetime}
\end{equation}

Eventually, if the fragmentation time happens before the observation time $t_\mathrm{obs}$, then we need to find the lengths of the two children loops assuming the loop fragmentation function \eqref{eq:lff-simple}.
The little sibling being drawn from a $-3/2$ power-law in $]\xi_c, \ell/2[$, its size $y$ is found using a uniform distribution $X \in \mathcal{U}(0, 1)$
\begin{equation}
    y = \qty[\xi^{-1/2} + X\xi^{-1/2} - X\qty(\ell/2) ^{-1/2}]^{-2} \, .
\end{equation}
For each of these daughter loops, we find their fragmentation time and continue recursively until $t_\mathrm{obs}$.

After a large number $k$ of fragmentation cascades, we are left with a population of loops at the observation time.
From this population, we can estimate the loop number density $\lnd(t_\mathrm{obs}, \ell)$ after accounting for the physical volume of the sampling space $V$.
Indeed, as we increase the number of fragmentation cascades $k$, we are effectively sampling a larger portion of space following
\begin{equation}
    \frac{C V}{3\nu - 3} \qty(t_\mathrm{obs}^{3\nu - 3} - t_\mathrm{min} ^{3\nu-3}) = k \, .
\end{equation}

This procedure is numerically intensive and does not yet include gravitational radiation.
Indeed, we have always assumed that the length of the loop remains constant, otherwise the fragmentation time would have another time dependence, which we could nonetheless be included in our expressions at the expense of solving a higher-order polynomial equation instead of a quadratic equation for \cref{eq:lifetime}.

This procedure presents a number of advantages. 
First, it is \emph{parallelizable}: One can draw each cascade independently on different machines and quickly obtain a very large number of samples. 
For instance in \cref{fig:comparison}, we obtain good convergence after $10^8$ samples in a couple of seconds on an 8-core CPU.
Additionally, the procedure is numerically stable by design.
As we increase the number of loops, we increase the statistics, but we do not accumulate errors.
And, since our loops keep fragmenting and accumulate on the smallest scales, we are guaranteed to have a lot of statistics and precision on small-scales.

Finally, let us note that the procedure is \emph{not manifestly scaling}.
As a consequence, it could be used to probe scenarios that break scaling, such as the impact of a constant correlation scale $\xi$ induced by the initial conditions of a Nambu-Goto simulations.
Therefore, if we see scaling appearing despite our fragmentation cascades, it means that scaling is a stable prediction.

\subsection{Integro-Differential Equation solver (IDE)}
\label{sec:ide}

Another approach consists in building a custom IDE solver taking advantage of the triangular nature of the master equation in scaling units \cref{eq:master-scaling}.
As we solve the Integro-Differential Equation, from $\gamma = \alpha$ down to $0$, we store the shape of the loop number density to compute the integrals in the range $]\gamma, \alpha[$ in the source terms.
This approach proves to converge faster than more traditional IDE solver such as Ref.~\cite{Karpel2018}.

In more details, our new IDE solver implements the following techniques.
First, it uses an \emph{implicit} scheme to discretize derivatives in \cref{eq:master} and \emph{adaptative steps}. 
The equation being stiff, an implicit scheme allows us to maintain good numerical stability. 

\begin{figure}
    \subfloat[Radiation era]{\includegraphics[width=.49\textwidth]{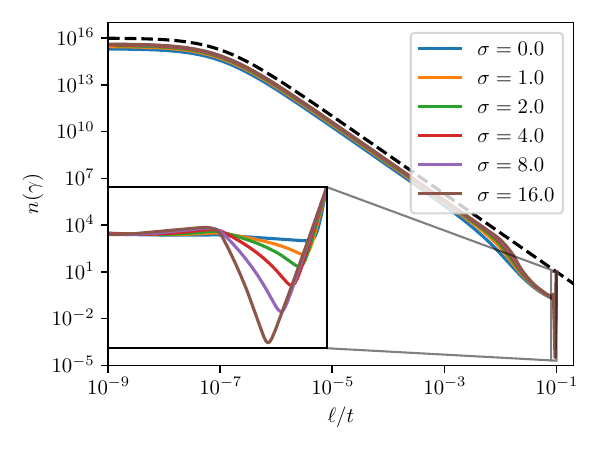}}
    \subfloat[Matter era]{\includegraphics[width=.49\textwidth]{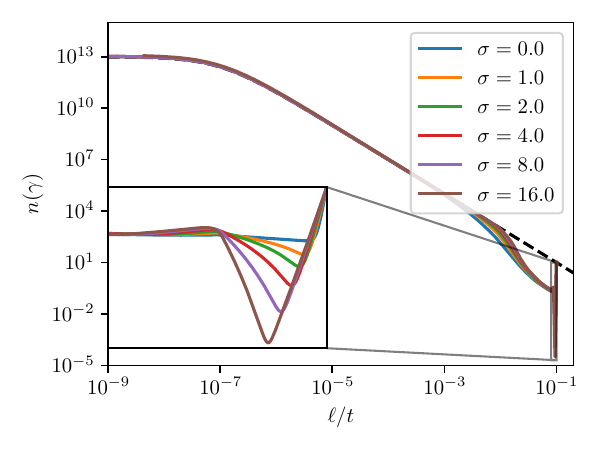}}
    \caption{Loop number density in scaling units for different values of the free parameter $\sigma$ encoding our uncertainties about the fragmentation model. 
    The zoom-in region shows how the value of $\sigma$ can help smoothing the sudden drop in the region $]\alpha - \xi_c, \alpha[$ and therefore ease the numerical resolution.
    The black dashed line corresponds to the one-scale model, assuming no fragmentation, i.e.\ $\lff = 0$.
    The parameters were set to $(C, \alpha, \chi, \nu, \xi_c) = (1, 0.1, 1/2, 10^{-2})$.}
    \label{fig:sigma_2}
\end{figure}

Second, as we solve for smaller and smaller values of $\gamma$, we store the value of the loop number density and use it to compute the integrals in the source terms.
For this purpose, we also use \emph{adaptative steps} to determine the resolution of our intermediate solution, and use a trapezoidal rule for the integrals.
Note that specific attention should be paid close to the boundary $\gamma$, see for instance the zoom-in region of \cref{fig:sigma_2}.
For this purpose, we integrate the rapidly varying $\lff$ analytically assuming $n(\gamma)$ constant in this neighbourhood for better numerical stability.

This IDE solver proves to be complementary to the ULM method.
Even though the IDE solver is not parallelizable, assumes complete scaling and does accumulate numerical errors down to the smallest scales, it has a number of advantages. 
Gravitational radiation, in the form of $\dot{\ell} = -\Gamma G \mu$ are taken into account and one could easily add other mechanisms for a loop to lose energy, for example through the emission of particles~\cite{Auclair:2019jip,Auclair:2021jud,Baeza-Ballesteros:2024otj}.
And contrary to the ULM method, we get direct access to the loop distribution, not a histogram.

Additionally, our IDE solver is more precise on the largest scales and is only expected to accumulate errors on the smallest scales.
Fortunately, we can use the ULM method to probe these smallest scales precisely and test the convergence of our results.
We tested the agreement between the two methods at length, and we illustrate one such test in \cref{fig:comparison}.

\begin{figure}
    \subfloat[Radiation era]{\includegraphics[width=.49\textwidth]{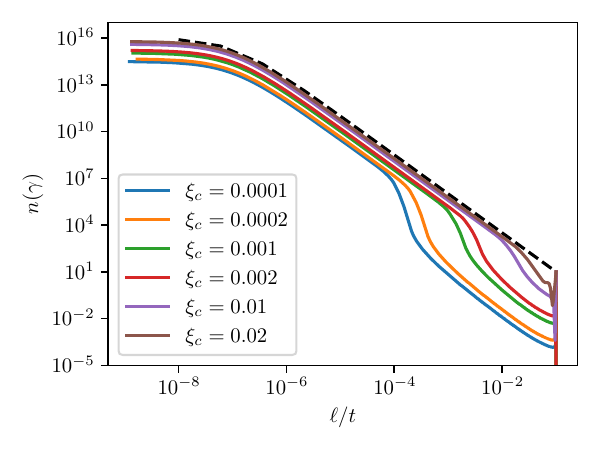}} %
    \subfloat[Matter era]{\includegraphics[width=.49\textwidth]{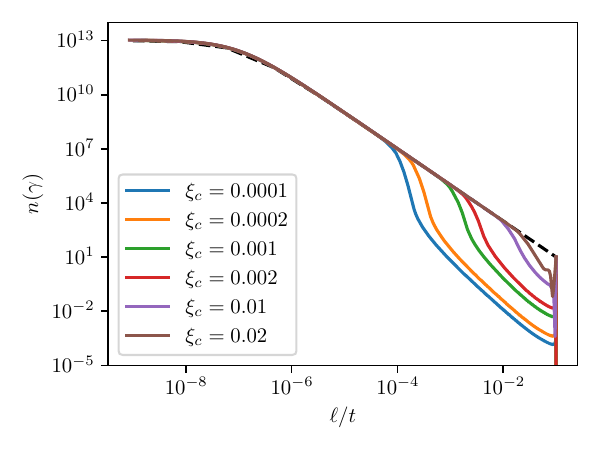}}
    \caption{
        Loop number density in scaling units $n(\gamma)$ for different values of the correlation length $\xi_c$ obtained with our IDE solver.
        The black dashed line corresponds to the one-scale model, assuming no fragmentation, i.e.\ $\lff = 0$.
        Parameters were set to $(C, \alpha, \chi, \Gamma G\mu, \sigma) = (1, 0.1, 0.2, 10^{-7}, 8)$.
    }
    \label{fig:xi_c}
\end{figure}

\begin{figure}
    \subfloat[Radiation era]{\includegraphics[width=.49\textwidth]{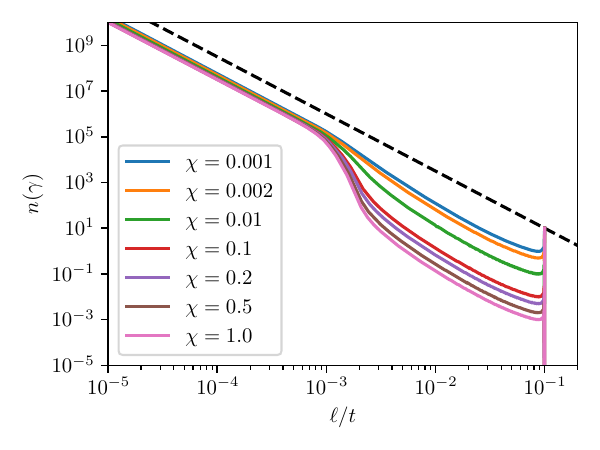}}
    \subfloat[Matter era]{\includegraphics[width=.49\textwidth]{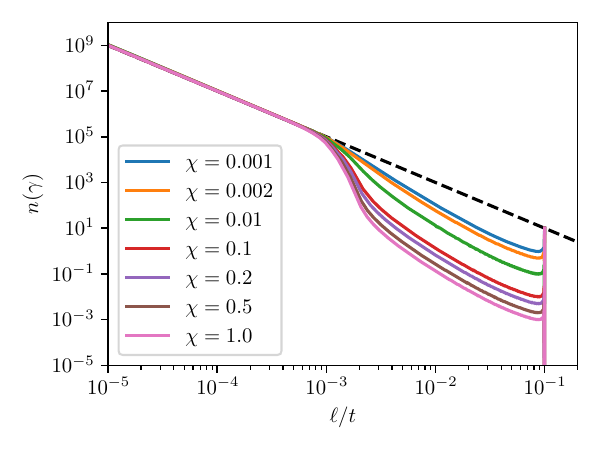}}
    \caption{Loop number density in scaling units for different values of the velocity $\chi$ encoding the strength of the fragmentation model. 
    The black dashed line corresponds to the one-scale model, assuming no fragmentation, i.e.\ $\lff = 0$.
    The parameters were set to $(C, \alpha, \nu, \Gamma G\mu, \xi_c) = (1, 0.1, 1/2, 10^{-7}, 10^{-3})$.}
    \label{fig:chi}
\end{figure}

\section{Results and Discussion}
\label{sec:discussion}

In this section, we present our findings on the loop number density including fragmentation.
We confirmed the convergence of our results by cross-correlating the ULM and IDE method, such as illustrated in \cref{fig:comparison}, and therefore we only present the results of our IDE solver in the following.
We illustrate our findings and the impact of the different parameters in \cref{fig:xi_c,fig:sigma_2,fig:chi}, compared to the so-called one-scale model which we define here as $\lff=0$, i.e.\ without fragmentations.
We intend to make our code public in the near future together with a database of precomputed loop number densities for a wide range of parameters.

As a general and important remark, the inclusion of fragmentation into small loops modifies the largest scales $\gamma \in ]\xi_c, \alpha[$ by reducing the amplitude of the loop number density and changing its shape from a pure $3\nu-4$ power-law to something resembling a $-5/2$ slope in both radiation and matter era.
On smaller scales, below the correlation length, the reduction in amplitude is less drastic in radiation era and totally absent in matter era. 
This latter behaviour proves to be a robust prediction of fragmentation cascades in all of our test cases.

Notably, we find that fragmentation does leave an imprint on the loop number density even for very low values of $\chi = 10^{-4}$ in \cref{fig:chi}.
This may indicate that $\chi$ is not a good parameter to assess whether fragmentation is relevant to the loop number density.
To build such a criterion, let us follow a loop produced at time $t_\star$ with size $\ell_\star = \alpha t_\star$. 
The probability that it survives long enough to reach the correlation length at time $\ell_\star / \xi_c$ is given by \cref{eq:g}
\begin{equation}
    G\qty(\frac{\ell_\star}{\xi_c}) = \exp[-\int_{\ell_\star / \alpha}^{\ell_\star / \xi_c} \tau^{-1} \dd{t}]
    \underset{\xi_c \ll \alpha}{\approx} \exp[-\frac{2\chi\alpha}{\xi_c^{5/2}}] \, .
\end{equation}
In the following, we assume to be in a regime in which fragmentation is unavoidable, that is assuming that $2\chi\alpha\xi_c^{-5/2} \gg 1$.

In this fragmentation regime, we find that the asymptotic behaviour of the loop number density can be reproduced, above and below the correlation length by the following scaling laws
\begin{equation}
    \eval{n(\gamma)}_\mathrm{rad} \propto 
    \begin{cases}
        C \alpha \chi^{-1} \xi_c^{3/2} \gamma^{-5/2} & \gamma \gg \xi_c \\
        C \alpha \chi^{-1/4} \xi_c^{3/4} (\gamma + \Gamma G\mu)^{-5/2} & \gamma \ll \xi_c \, .
        \label{eq:scaling-rad}
    \end{cases}
\end{equation}
Note that this formula is not continuous about $\xi_c$.
Indeed, as shown on \cref{fig:sigma} the transition between above and below the correlation length can be very sharp.
One could be intrigued by the limit $\chi \to 0$ which seems to indicate that the loop number density diverges when we turn fragmentation off.
But one should bear in mind that we are working in the limit in which fragmentation is unavoidable, ie $2\chi\alpha\xi_c^{-5/2} \gg 1$.
Hence, the limit $\chi \to 0$ is outside the regime of validity of this equation if all the other parameters remain fixed.

For the matter era, we find a different set of asymptotic behaviours summarized in the formulas below
\begin{equation}
    \eval{n(\gamma)}_\mathrm{mat}\propto
    \begin{cases}
        C \alpha \chi^{-1} \xi_c^{3/2} \gamma^{-5/2} & \gamma \gg \xi_c \\
        C \alpha (\gamma + \Gamma G\mu)^{-2} & \gamma \ll \xi_c\, .
        \label{eq:scaling-mat}
    \end{cases}
\end{equation}
This result is interesting for two separate reasons. 
First, the loop number density on small scales does not seem to be affected at all by the occurrence of fragmentation.
This seems to suggest that observations probing a late-time Universe filled with cosmic strings may yield more robust predictions, insensitive to fragmentation.
Second, the loop number density above the correlation length, while not being a perfect power-law, is close to a $-5/2$ slope compared to the prediction of a $-2$ slope in the one-scale model.
This is all the more exciting since it could explain while different numerical simulations of cosmic strings have found steeper slopes for the loop number density in matter era~\cite{Ringeval:2005kr,Blanco-Pillado:2019tbi}.

\begin{figure}
    \subfloat[Radiation era]{\includegraphics[width=0.49\textwidth]{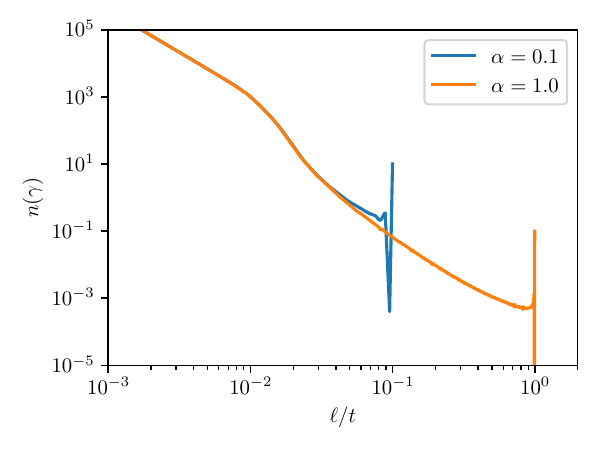}} %
    \subfloat[Matter era]{\includegraphics[width=0.49\textwidth]{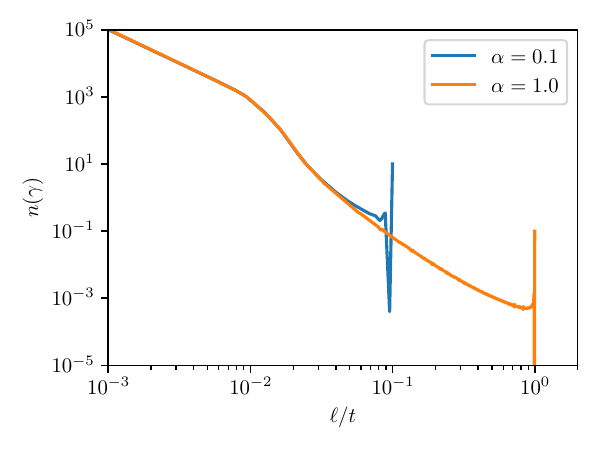}}
    \caption{
        Loop number density in scaling units $n(\gamma)$ obtained by moving the boundary condition, i.e.\ the position of the loop production scale $\alpha$.
        We rescaled the value of $C$ to keep $C \alpha$ constant.
        Parameters were set to $(\chi, \Gamma G\mu, \xi_c, \sigma) = (0.2, 10^{-7}, 10^{-2}, 8)$.
    }
    \label{fig:alpha}
\end{figure}

Finally, one could raise doubts about the validity of our whole framework. 
Indeed, we have traded the non-linear rate equation defined on all scales, for a linear and collision-less model defined on sub-Hubble scales.
Therefore, one may legitimately ask about the impact of our new boundary at $\gamma = \alpha$.
Perhaps the most spectacular impact of the boundary, is the sudden drop for $\gamma \lesssim \alpha$ highlighted in \cref{fig:sigma_2}.
This drop is an obvious artefact of our method and does not carry physical meaning.
Nonetheless, it is of primordial importance to resolve it numerically to ensure the global convergence of our IDE solver.
Fortunately, as our free parameter $\sigma$ goes to $0$, it tends to smoothen this sudden drop as well as the transition occurring around the correlation length $\xi_c$ without impacting significantly the rest of the loop number density. 
This property of $\sigma$ helps us achieve better numerical stability as $\sigma$ gets closer to zero and increases our faith in the robustness of our result.

Regarding the dependence on the location of the boundary $\alpha$, we show in \cref{fig:alpha} that, apart from the boundary artefact discussed above, changing the location of the boundary does not affect the loop number density, as long as the product $C \alpha$ remains constant.
Even though the parameters $C$ and $\alpha$ have distinct physical interpretations, it is remarkable that such a simple and universal rule exists for both radiation and matter era.
This property appears directly in the asymptotic expressions of \cref{eq:scaling-rad,eq:scaling-mat} 

\section{Conclusion}

Inspired by the three-scale model and in particular by the rate equation of Refs.~\cite{Steer:1998qg,Magueijo:1999qc}, we built a new model to probe the impact of loop fragmentation on the loop number density in both radiation and matter era.
To solve it, we presented two robust numerical methods: the Unconnected Loop Model (ULM) -- which samples random fragmentation cascades -- and a custom Integro-Differential Equation (IDE) solver taking advantage of the specific ``triangular'' property of our rate equation.
Both these methods benefit from the absence of loop collisions in our model.

We applied these new techniques to a simplified model of fragmentation, which assumes that cosmic strings behave as random walks above a scaling correlation length $\xi = \xi_c t$.
The behaviour below the correlation being characterized by a single free parameter $\sigma$.
We find that loop fragmentation can have a significant impact on the large scales, even for small values of the velocity parameter $\chi$.
Fragmentation is characterized by a decrease in amplitude and a deviation from a pure power-law, around a slope of about $-5/2$ in both radiation and matter era.

We find that cosmic string loops during radiation era are perhaps the most affected by fragmentation.
On the small scales $\gamma \ll \xi_c$, of relevance for gravitational waves observations, fragmentation reduces substantially the loop distribution by a factor of order $\sqrt{\alpha} \xi_c^{3/2} / \chi$ compared to the one-scale model, and with a dependency on the parameter $\sigma$ encoding our ignorance of the details of loop fragmentation.
This implies that GW experiments at high frequencies, such as ground-based experiments probing the radiation era, are sensitive to the details of loop fragmentation.

To the contrary during matter era, the small scales are not sensitive to loop fragmentation, for all parameters.
This seems to indicate that GW experiments at low frequencies, such as pulsar timing arrays probing the matter era, are more robust against the details of loop fragmentation.

It is certainly premature to make observational predictions at this stage but we can still highlights the key effects that fragmentation may have on GW observations.
First, the amplitude of the GW power spectrum at a given frequency is mostly determined by the amplitude of the loop number density at a given time in the past at the scale $\gamma = \Gamma G\mu$ (see for instance Refs.~\cite{Sousa:2020sxs,Auclair:2020oww}), that is inaccessible to numerical simulations.
All current cosmic string models have to interpolate between simulations of the largest scales and this small gravitational wave scale.
We have shown that accounting for fragmentation may alter this correspondance, and therefore the overall amplitude of the GW power spectrum.
Second, we have shown that fragmentation affects differently the radiation and matter-dominated eras, and consequently the low and high frequency parts of the power spectrum.
Therefore, we may expect the overall shape of the GW power spectrum to be modified, particularly hierarchy between the high-frequency plateau and the matter-era bump, as for example the models C-1 and C-2 of Ref.~\cite{LIGOScientific:2021nrg}.

Our model for loop fragmentation is still rather simplistic but shows that loop fragmentation deserves to be studied in more depth, especially in the context of the future generation of ground-based detectors which will study the Stochastic Background of GWs from cosmic strings.
This might also be of interest for LISA, which will be able to probe the Stochastic Background around the millihertz, and therefore to probe the transition between radiation and matter dominated eras.
In this context, we plan to calibrate the loop fragmentation function with new Nambu-Goto numerical simulations in a later publication.
Notably, the Unconnected Loop Model opens the possibility to calibrate directly our fragmentation cascades against the fragmentation graphs found in Nambu-Goto simulations.

\acknowledgments

P.A.\ thanks Christophe Ringeval and Danièle Steer for their ongoing encouragement to complete this work.
P.A.\ also thanks Jorge Baeza-Ballesteros, Ed Copeland, Daniel Figueroa, Mark Hindmarsh and Tanmay Vachaspati for useful discussions during the preparation of this manuscript.
P.A.\ would like to thank Nordita for support and hospitality during the program ``Numerical Simulations of Early Universe Sources of Gravitational Waves'' (July-August 2025), Stockholm.
Pierre Auclair is a Postdoctoral Reseacher of the Fonds de la Recherche Scientifique – FNRS.

\appendix

\section{ULM equations with a $\sigma-$regulator}
\label{app:sigma}

We introduce a regulator $\sigma$ on the fragmentation function on small scales
\begin{equation}
    \lff(y, \ell - y; \ell) = \frac{\chi \ell}{(\xi y)^{3/2}} \Theta(y - \xi) + \frac{\chi \ell}{\xi^3} \left(\frac{y}{\xi}\right)^{\sigma} \Theta(\xi - y) \, .
\end{equation}
With this regulator, the function becomes continuous at $y = \xi$.
This property helps with the numerical stability and allows us to marginalize over our uncertainties below the fragmentation scales.
The formulas in the main text need to be adapted for this regulator.

The regulator $\sigma$ does not change the formation time of loops from the loop production function, but impacts its lifetime.
As such, the rate of fragmentation now reads
\begin{equation}
    \tau^{-1} 
    = \int_0^{\ell / 2} \lff(y, \ell) \dd{y}
    = \begin{cases}
        \displaystyle \frac{\chi \ell}{(1 + \sigma)(\xi_c t)^2} \left(\frac{\ell}{2 \xi_c t}\right)^{1 + \sigma} 
        & \text{if } \ell < 2 \xi_c t \\
        \displaystyle \frac{2 \chi \ell}{(\xi_c t)^2} - \frac{2 \sqrt{2 \ell} \chi}{(\xi_c t)^{3/2}} + \frac{\chi \ell}{(1 + \sigma)(\xi_c t)^2} 
        & \text{if } 2 \xi_c t < \ell
    \end{cases}
\end{equation}
Therefore, the rate of fragmentation does not vanish when the loop length goes below the fragmentation scale anymore and the death time can be arbitrary large.

We obtain the life expectancy by integrating the rate with respect to time and taking the exponential.
Assuming that the loop was formed during the ``fragmentation'' period, that is $2 \xi_c t_\star < \ell$
\begin{equation}
    \int_{t_\star}^t \tau^{-1} \dd{t'}
    = \begin{cases}
        \displaystyle \frac{2\chi \ell}{\xi_c^2} \left(\frac{2\sigma + 3}{2\sigma + 2}\right) \left(t_\star^{-1} - t^{-1}\right) 
        - \frac{4 \chi \sqrt{2\ell}}{\xi_c^{3/2}} \left(t_\star^{-1/2} - t^{-1/2}\right)
        & \text{ if } 2 \xi_c t < \ell \\
        \displaystyle \int_{t_\star}^{\ell / (2 \xi_c)} \tau^{-1} \dd{t'} 
        + \frac{\chi \ell}{(1+\sigma) \xi_c^2} \left(\frac{\ell}{2\xi_c}\right)^{1 + \sigma} \frac{1}{2 + \sigma}\left[\left(\frac{\ell}{2 \xi_c}\right)^{-2-\sigma} - t^{-2-\sigma}\right]
        & \text{ if } \ell < 2 \xi_c t \\
    \end{cases}
\end{equation}
Otherwise, if the loop is formed after its ``fragmentation period''
\begin{equation}
    \int_{t_\star}^t \tau^{-1} \dd{t'}
    = \frac{\chi \ell}{(1+\sigma) \xi_c^2} \left(\frac{\ell}{2\xi_c}\right)^{1 + \sigma} \frac{1}{2 + \sigma}\left[t_\star^{-2-\sigma} - t^{-2-\sigma}\right] \,.
\end{equation}
Then, one needs to draw $X \in \mathcal{U}(0, 1)$ and find the death time satisfying
\begin{equation}
    \exp\left(-\int_{t_\star}^t \tau^{-1} \dd{t'}\right) = X  \,.
\end{equation}
Note that it suffices to check $\ln X$ against
\begin{equation*}
    - \int_{t_\star}^{\ell / (2 \xi_c)} \tau^{-1} \dd{t'} 
\end{equation*}
to know which of the two branches one should use to find the solution.
Notice that the introduction of the $\sigma$-regulator does not mean that loops are never stable!

Eventually, we are also interested in the lengths of the child loops, which can now be arbitrarily small.
First, we need to integrate the fragmentation function with $x < \ell / 2$
\begin{equation}
    \int_0^{x} \lff(y, \ell - y; \ell) \dd{y}
    = \begin{cases}
        \displaystyle \frac{\chi \ell}{(1 + \sigma)\xi^2} \left(\frac{x}{\xi}\right)^{1 + \sigma}
        & \text{ if } x < \xi \\
        \displaystyle
        \frac{\chi \ell}{(1 + \sigma)\xi^2}
        + \frac{2 \chi \ell}{\xi^{3/2}} \left(\xi^{-1/2} - x^{-1/2}\right)
        & \text{ if } \xi < x
    \end{cases}
\end{equation}
To get the size of the fragmented loop, we draw $X \in \mathcal{U}(0, 1)$ and then solve for 
\begin{equation}
    \int_0^{x} \lff(y, \ell - y; \ell) \dd{y} = X \int_0^{\ell / 2} \lff(y, \ell - y; \ell) \dd{y} \, .
\end{equation}

\bibliographystyle{JHEP}
\bibliography{biblio}

\end{document}